\documentclass[preprint,showpacs,preprintnumbers,amsmath,amssymb]{revtex4}

\usepackage{graphicx}
\usepackage{dcolumn}
\usepackage{bm}
\usepackage{color}

\begin{document}

\title{Specific Heat Studies of Pure Nb$_3$Sn
Single Crystals at Low Temperature}

\author{R. Escudero}\email[Author to whom correspondence should be
addressed. Email address:]
{escu@servidor.unam.mx}
\author{ F. Morales}
\affiliation{Instituto de Investigaciones en Materiales, Universidad
Nacional Aut\'{o}noma de M\'{e}xico. A. Postal 70-360. M\'{e}xico, D.F.
04510 MEXICO.}

\author{S. Bern\`{e}s}
\affiliation{Universidad Aut\'{o}noma de Nuevo Le\'{o}n, Facultad de
Ciencias Qu\'{\i}micas. Monterrey Nuevo Le\'{o}n, MEXICO.}

\date{\today}

\begin{abstract}
Specific heat measurements performed on high purity vapor-grown
Nb$_3$Sn crystals show clear features related to both the martensitic
and superconducting transitions. Our measurements indicate that the
martensitic anomaly does not display hysteresis, meaning that
the martensitic transition could be a weak first or a second order
thermodynamic transition. Careful measurements of the two transition
temperatures display an inverse correlation between both
temperatures. At low temperature specific heat measurements show
the existence of a single  superconducting
energy gap feature.
\end{abstract}

\pacs{ }

\maketitle

\section{Introduction}

Nb$_3$Sn is a well known intermetallic compound with the cubic A15
structure at room temperature. It displays two interesting features
at low temperature, a cubic-to-tetragonal martensitic transformation
in the range $T_M$ = 40-50 K, and a superconducting transition at
about $T_C$ = 18 K \cite{mail67,mail67a,Weger71}. The martensitic
transition has generated a great deal of interest but it is still not
completely understood. Theoretical and experimental results are not
conclusive  about the order of the martensitic anomaly. Early
specific heat studies did not give clear evidence about the
thermodynamic order of the transition, nor if there is a correlation
between the martensitic and superconducting transitions temperatures.
In addition, recently specific heat measurements in the
superconducting state of Nb$_3$Sn samples prepared by solid state
diffusion reaction, have been interpreted as showing the presence of
an intrinsic second superconducting gap,  affecting the electronic
density of states \cite{guri04}.

One reason for this lack of understanding of the cubic-tetragonal
transformation is that this thermodynamic anomaly was not clearly
observed in early  heat-pulse calorimetry studies. This may be due to
noise in the data \cite{vieland69}, anomaly too small in the total
specific heat in ac calorimetry \cite{chu74}, or the absence of the
anomaly \cite{guri04}. It has been inferred, however, that the cubic
to  tetragonal transformation must be a first order thermodynamic
transition, with a proposed Jahn-Teller mechanism \cite{vieland71,
labbe66} which should be observable as hysteresis in specific heat
measurements through the transition, and in entropy measurements. It
is very important to mention at this point that a martensitic
transition must be necessarily a first order thermodynamic
transition, showing a hysteretic behavior in the specific heat and in
resistivity-temperature measurements \cite{Tsumura88, Makita}. In
fact in materials displaying this kind of transformations, as the
Al-Cu-Zn alloy, which can be considered as the  prototype martensitic
alloy \cite{Lovey},  where a phase transformation is observed from an ordered
b.c.c ($\beta_1$) parent phase  to an ordered {\bf 18R} phase,  at
different temperatures depending on composition \cite{Tsumura88,
Lovey}. In specific heat measurements one can observe the typical
hysteresis of the  thermodynamic transition at two different
temperatures, as well in transport measurements. For the Nb$_3$Sn
case this hysteresis has not been observed  with enough precision.
Among the reasons for the lack of observation of this anomaly, could
be non-optimal characteristics of the samples, internal stress,
samples with incorrect stoichiometry, compositional inhomogeneity, or
inclusive technical deficiencies of the experimental equipment when
performing the specific heat measurements \cite{testardi75}. It is
important to mention that frequently this A15 alloy, presents
impurities due to formation of other compositions and inclusive Sn
excess has been often observed. There are two important characteristics
about this alloy that must be understood. The first one
 is related to the debate about the order of the high
temperature transition; the so called martensitic anomaly. The other
is about the influence of the martensitic transition on the
electronic density of states, and therefore on the superconducting
transition temperature \cite{weber82}. Respect to the first point,
Labbe and Friedel \cite{labbe66} in their studies related to  the
cubic to tetragonal transition  proposed that it may involve a
Jahn-Teller distortion, occurring in the crystalline structure by the
effect of the one-dimensional Nb chains. These chains are one of the
crystallographic features of the A15 compounds. From this point of
view calculations based on electron sub-bands models have predicted
that the cubic to tetragonal transition is first order. However,
experimentally Vieland and co-workers \cite{vieland71} presented only
indirect evidence that the structural transition is first order
(indirect in the sense that the thermodynamic order was inferred from
non-thermodynamic measurements), as predicted by theories invoking a
band Jahn-Teller distortion. Nevertheless, other experimental results
have shown an absence of latent heat and hysteresis at the transition
temperature, which  has been interpreted as evidence for a
thermodynamic transition of the second order \cite{vieland69}.
Lastly, we must mention that another physical possibility of the
martensitic transformation in this compound could  be related to a
Peierls distortion, particularly in this case, of the formation of a
Charge Density Wave (CDW) promoted by the infinitum Nb chains, as was
theoretically formulated by Gorkov, Bhatt and McMillan.

The aim of this work is two fold; clarify the role of the so called
martensitic transition on the superconducting temperature, and to
prove the possible existence of  two superconducting energy gaps, as
recently claimed. The results of this investigation can be
resumed in the following: Thermodynamics characteristics of  seven
Nb$_3$Sn single crystals were determined  by mean of specific heat
measurements. The results show the clear martensitic anomaly around
50 K, and this does not display hysteretic behavior, as expected for
a first order thermodynamic transition. Thus, suggesting a weak first
order or a second order transition.

We observe  a clear correlation between the superconducting and the
so called martensitic transition temperatures. In addition we
observed in the  specific  heat  measurements at low temperatures the
existance of only a feature related with the  presence of only a
single energy gap.

In addition, in order to  corroborate that our results are obtained
in good specimens, we performed  X-ray diffraction analysis which
show that the samples are single crystals.

\section{Experimental Details}

The Nb$_3$Sn samples  were grown over a period of four months by
closed-tube vapor transport with iodine vapor as the transport agent.
From the batch of samples were selected seven single crystalline
specimens for our measurements, they consist of just one to several
crystals with mm-size. Crystallographic characteristics where
examined at room temperature by X-ray diffraction, using a Siemens P4
diffractometer equipped with Mo-K$\alpha$ radiation
($\lambda$=0.71073 \AA).

Electrical resistivity was determined by the four probe method. These
samples were found to have resistivity ratio $R$(300 K)/$R(T_C)$ = 18
and an extrapolated residual resistivity ratio of $R$(300 K)/$R$(0 K)
= 50, indicating the high degree of purity and crystalline perfection
\cite{arko78}. Early studies with these crystals were performed up to
18 Tesla as reported by  Stewart, Cort and Webb \cite{stewart81}. As
an antecedent of the quality of the crystals, some of the oriented
single crystals from this same batch were used for dHvA oscillation
studies of the Fermi surface \cite{arko78}. The  crystals were
measured and characterized by magnetization versus temperature,
resistance versus temperature, and specific heat versus temperature
in order to study the characteristics of the so-called martensitic
transformation and the superconducting transition temperature.

\begin{table}
\caption{Crystallographic data for two single crystals NS1 and NS4.}
\begin{tabular}{lll}
  \toprule
  Compound & NS1 (5.31 mg) & NS4 (9.5 mg) \\    \colrule
  Empirical formula & Nb$_3$Sn & Nb$_3$Sn \\
  Formula weight & 397.42 & 397.42 \\
  Color, habit & Metallic, irregular\ &  Metallic, prism \\
  Crystal size [mm] & 0.3 $\times$ 0.2 $\times$ 0.2 & 0.4 $\times$ 0.4 $\times$ 0.4 \\
  Space group &\emph{P m-3n}  &\emph{P m-3n} \\
  \emph{a} [\AA]   &5.2700(9)   &5.2531(13) \\
  \emph{V} [\AA$^3$]  &146.36(4)  & 144.96(6) \\
  \emph{Z}   &2   &2 \\
  $\rho_{calcd}$ [g cm$^{-3}$] &9.018   &9.105 \\
  $\mu$[mm$^{-1}$]    &19.54   &19.73 \\
2$\theta$  Range [$^{\circ}$]    &11 - 70 &11 - 70 \\
Reflections collected   &2147   & 1958 \\
$<I/\sigma (I)>$   & 53  &109 \\
Independent reflections (R$_{int}$)  &72 (0.1978) &69 (0.1953) \\
Completeness [\%]    &97.2    &95.8 \\
Transmission factors [min., max.]   &0.015, 0.348   & 0.014, 0.347 \\
Final $R$ indices (all data) $R_1$, $wR_2$  &0.054, 0.105    &0.062, 0.167 \\
Goodness-of-fit on $F^2$   &1.399   &1.309 \\
Extinction coefficient  &0.82(18)    &0.8(2) \\
Data / restraints / parameters  & 72 / 0 / 5 &69 / 0 / 5 \\
Largest difference peak / hole [e. \AA$^{-3}$] & 4.369, -1.789 & 4.238, -2.807 \\
  \botrule
  \textbf{Final Geometric Parameters}& &\\ \hline
  Distance& & \\ \hline
  Nb-Sn (\AA)  & 2.9460 (5) & 2.9366 (7) \\
Nb-Nb (\AA)  & 2.6350 (5)  &2.6266 (6) \\
Nb...Nb (\AA) &3.2272 (5)  &3.2169 (8) \\ \botrule
\end{tabular}
\label{table1}
\end{table}

Specific heat measurements were performed between room temperature
and 2 K under magnetic field below 0.1 Oe, using a thermal relaxation
technique with a Quantum Design PPMS calorimeter. Calibrated addenda
corrections for the sample holder were subtracted, thus the specific
heat measurements show the corrected values for all samples. In the
range from 2-60 K measurements were taken in multiple cooling and
warming cycles.

\begin{figure}[btp]
\begin{center}
\includegraphics[scale=0.7]{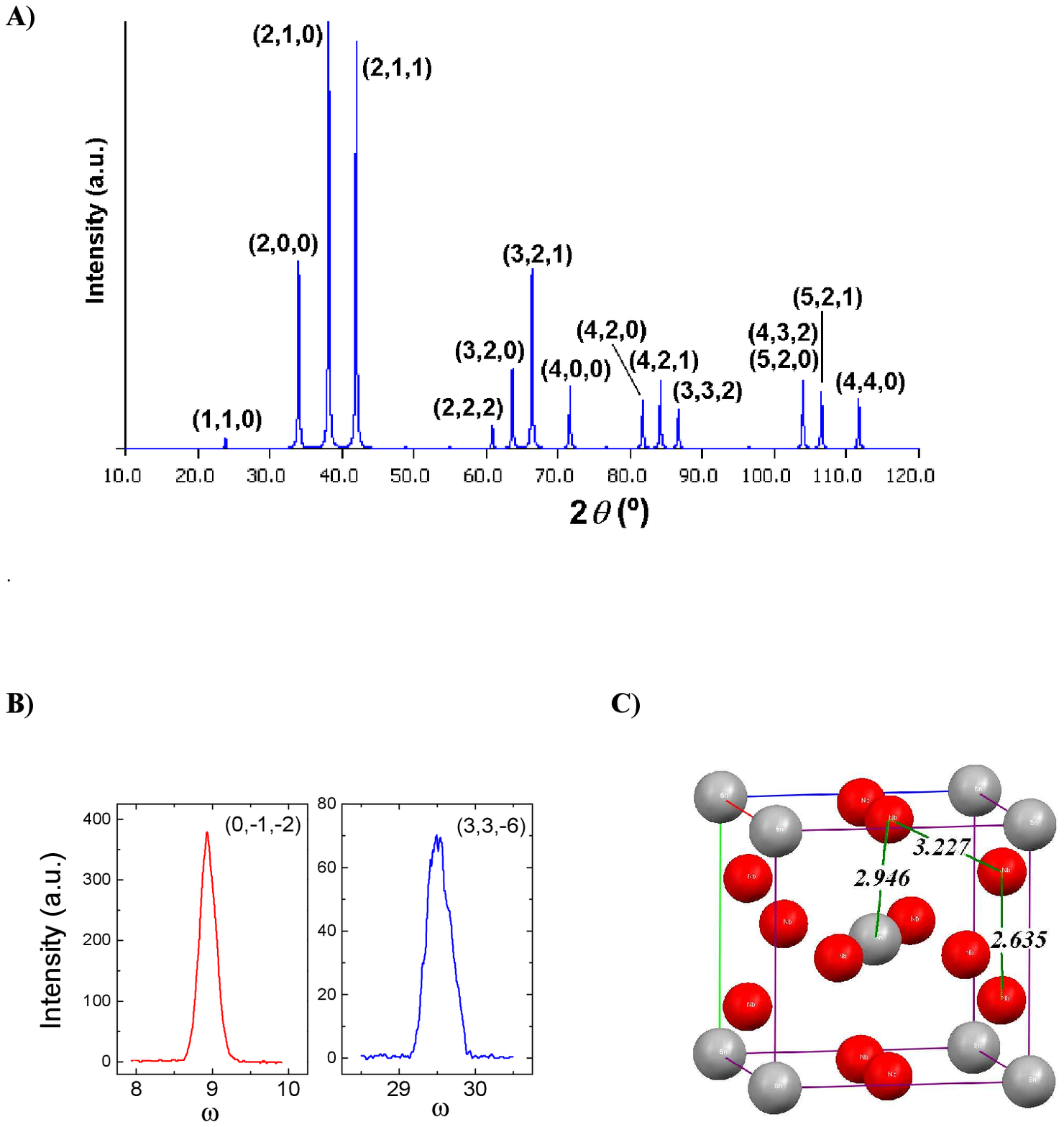}
\end{center}
\caption{(Color online) A). Calculated powder diffraction pattern for
Nb$_3$Sn ($\lambda$ = 1.5405 \AA, no preferred orientation), computed using
X-ray data of the single crystal NS1. Software: CaRine Crystallography
(Release 3.1). C. Boudias \& D. Monceau, 1998. B). Scans for reflections
(0,-1,-2) and (3,3,-6) for crystal NS1 \cite{siemens}.
C). Crystalline structure of Nb$_3$Sn. Distances  in \AA, grey and
red spheres correspond to Sn and Nb respectively.}
\label{fig1}
\end{figure}

\section{Results and Discussion}

For the crystalographic characterization we studied two typical
samples by X-ray diffraction, hereafter called  NS1 and NS4, with
mass about 5.31 and 9.5 mg, respectively. X-ray characteristics were
taken at a temperature of  298(1) K using a Siemens P4 diffractometer
equipped with Mo-K$\alpha$ radiation($\lambda$ =0.71073 \AA).
Although samples are large and display somewhat irregular shapes,
they revealed to be single crystals, as reflected by well defined,
symmetric Bragg diffraction peaks, see  panel A of Fig. \ref{fig1}.
This Fig. shows the refined data obtained by computing generated
spectrum of the diffraction pattern of the single crystal NS1,
whereas the panel B shows two reflections at two particular positions
in the the reciprocal space. In this lower panel $\omega$ corresponds
to the Bragg angle which according to the diffractometer geometry
corresponds to $2\theta$/2. The  two reflections, one at low angle
$\omega = 9^\circ$, the another at high angle at $\omega =
29.6^\circ$. are representative features for a  good single crystal.
The reflection peaks at low angles necessarily needs to have  four
characteristics: a single well defined maximum, the peak must be
highly symmetric, a ratio signal to noise high, and the same amount
of noise at the left and right of the peak. The reflection at
(0,-1,-2) presents those characteristics. Whereas the peaks at high
angle must have preferentially,  the higher ratio signal to noise,
they can be non-symmetric because the monochromaticity of the
diffraction apparatus need to be taken into account due to the two
contribution of the X-ray diffraction beam (K$_{\alpha I}$ +
K$_{\alpha II}$). Thus, accordingly the ratio signal to noise is
necessarily low due to the fact that the atoms scattering factor
decreases as the Bragg angle increases. This is intrinsic to any
crystal, good or with low quality. In the NS1 crystal, the reflection
at (3,3,-6) show clearly the peak well defined, the noise is constant
at the two ends of the peak, this  meaning  that the crystal presents
an ideal crystallinity. In addition, we remark that no secondary
diffraction patterns were observed for possible impurities, neither
diffuse scattering. For each sample, a complete diffraction sphere
was collected \cite{siemens} at the highest available resolution
(0.62 \AA, 2$\theta_{max}$ = 70$^\circ$). As expected, crystals
belong to space group Pm-3n and the structure of Nb$_3$Sn is an
A15-type arrangement, as previously described \cite{testardi75}.
Atomic positions were refined \cite{sheldrick} on the basis of
absorption-corrected data \cite{walker}. A characteristic parameter
in a crystal is the  high value of the extinction coefficient, which
converges to identical values for both samples: so in those crystals
we found  0.82(18) for NS1 and 0.8(2) for NS4 \cite{sheldrick}.
Assuming that applied correction covers mixed primary and secondary
extinctions, this result suggests that samples should have similar
block sizes and similar concentrations of randomly distributed
dislocations \cite{masimov}.

In Table 1 we present many of the characteristics of two specimens
measured. Both samples are characterized by rather short unit cell
parameters, \emph{a} = 5.2700(9) and \emph{a} = 5.2531(13) \AA, for
NS1 and NS4, respectively, while the accepted value found in the
literature for crystalline Nb$_3$Sn is \emph{a} = 5.29 \AA\
\cite{pdf}. Interestingly, NS1 and NS4 have significantly different
cell parameters, and, as a consequence, cell volume is reduced by ca.
1\% in NS4, compared to NS1. Calculated densities present  the same
1\% drop. However, using diffraction data, a confident interpretation
of such a cell contraction in terms of intrinsic vacancies in the
alloy cannot be carried out, at least if departures from Nb$_3$Sn
stoichiometry remain small. In contrast, the high resolution of
diffraction data allows to accurately determine distances in the
solid (Table 1). The shortest Nb...Nb separation is reduced from
2.6350(5) \AA\ in NS1 to 2.6266(6) \AA\ in NS4. In the same way,
Nb...Sn separations in NS1 and NS4 are 2.9460(5) and 2.9366(7) \AA,
respectively.

The two samples examined by X-ray included in this work, NS1 and NS4,
have two different characteristics in the high temperature anomaly,
see Fig 2. Whereas NS1 has a very well defined sharp peak at the
specific heat characteristic, which is starting at about 50 K, NS4
presents only a small feature  starting at about 46 K.  A possible
interpretation about the sharpness of these two anomalies may be
related to the crystalline structure; if Nb vacancies are present in
the structure, then the cell parameters and volume will be reduced,
as observed in samples NS1  and NS4. We may speculate that the
implication of this behavior might be related to a Peierls distortion
in the Nb chains \cite{gorkov73, bhatt76}. Thus, the charge density
wave (CDW) created due to the distortion will open an energy gap in
the direction of the chains, and possible better formed if the chains
are without Nb vacancies, that if there exist deficiencies or
vacancies. Therefore, this will reduce the size and sharpness of the
anomaly.

We also can not discard another possibility; this may be
due to the  crystal inhomogeneities. As crystals start to grow, it is
not surprising to observe sample to sample variations in the A15
structure, albeit differences comparing NS1 and NS4 samples which are
considered small. As NS4 is a crystal with a bulk volume about two
times that NS1, inhomogeneities may be seen as a natural consequence.
However in our set of samples the slow growing process performed over
a period of four months, the inhomogeneities may be very small and
our first speculation related to CDW nesting can not be discarded. A
last important evidence of the perfection of these crystals, is
supported by studies in the same batch of samples by Arko et al.,
\cite{arko78} related to the Fermi surface with experiments of the de
Haas-van Alphen effect, which agree considerable well with
theoretical studies. In the de Haas-van Alphen experiments an
important and necessary aspect is related to the inhomogeneities on
the crystal structure under study. Those inhomogeneities  must be as
small as possible in order to observe features of the Fermi surface.

\subsection{Specific Heat Measurements}

\begin{figure}[btp]
\begin{center}
\includegraphics[scale=0.8]{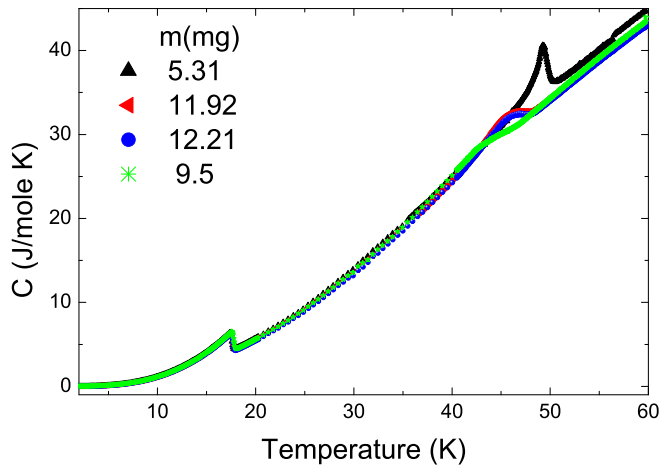}
\end{center}
\caption{(Color online) Specific heat vs.
temperature of four Nb$_3$Sn single crystal measured between 2
and 60 K. Clearly it is observed both transitions, the high temperature
transition shows the martensitic anomaly in four samples, the anomaly
in the samples presents different shapes and sizes, no extra anomaly was
observed below the superconducting transition.}
\label{fig2}
\end{figure}

The specific heat measurements were performed as before was
explained. In Fig. \ref{fig2} we show data of four Nb$_3$Sn samples
measured in the interval from 2 to 60 K.  An interesting
characteristic in those measurements is the notable different shapes
and temperatures of the high temperature anomaly in the specimens.
Whereas  the superconducting transition temperature at about 18 K
almost presents only a small variation in size and temperature.

\begin{figure}[btp]
\begin{center}
\includegraphics[scale=0.8]{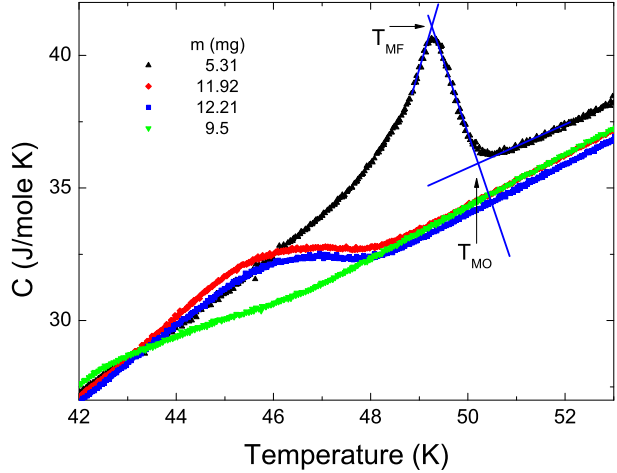}
\end{center}
\caption{(Color online) The specific heat \emph{vs}.
temperature of four Nb$_3$Sn single crystal between 42 K and 53 K with
warming and cooling cycles. Note that there is no observable
hysteresis around the martensitic transition. The straight lines in
one of the curves is presented to show the form to perform the
determination of the transition onsets and final temperatures. }
\label{fig3}
\end{figure}

\begin{figure}[btp]
\begin{center}
\includegraphics[scale=0.8]{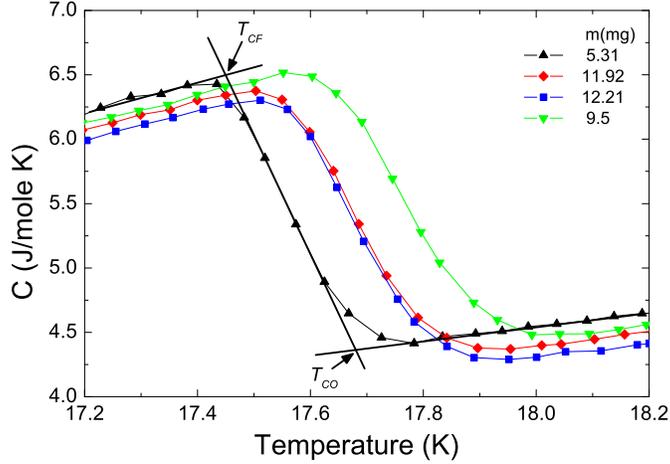}
\end{center}
\caption{(Color online) Specific heat vs. temperature of the four
samples shown in figure 2 around the  superconducting transition
temperature. The straight lines show the form to determine the
transition onsets and finals temperatures.}
\label{fig4}
\end{figure}

Fig. \ref{fig3} depicts the specific heat of the four samples of
Nb$_3$Sn crystals, the martensitic anomaly ($T_M$) occurs in the
interval from 42 to 53 K, the martensitic anomaly presents different
sizes. In this plot the curves were measured in  cooling and warming
cycles, there are no hysteresis at all; thus suggesting a weak first
or second order thermodynamic transition. Some of these curves
display a much clearer anomaly at $T_M$ than has been observed before
\cite{vieland69,chu74}. In Fig. \ref{fig4} the superconducting
transitions of the same samples of Fig. \ref{fig3} are displayed. The
superconducting transition temperature onsets  have a small but
discernible differences between the four samples, with the high onset
for the $9.5$ mg sample and the minimum for the $5.31$ mg sample. The
manner that we used for the determination of the transition
temperatures; onsets and finals, is also show in   Fig. \ref{fig4}.
There the straight lines intersections indicate the transition
temperatures.

\begin{figure}[btp]
\begin{center}
\includegraphics[scale=0.8]{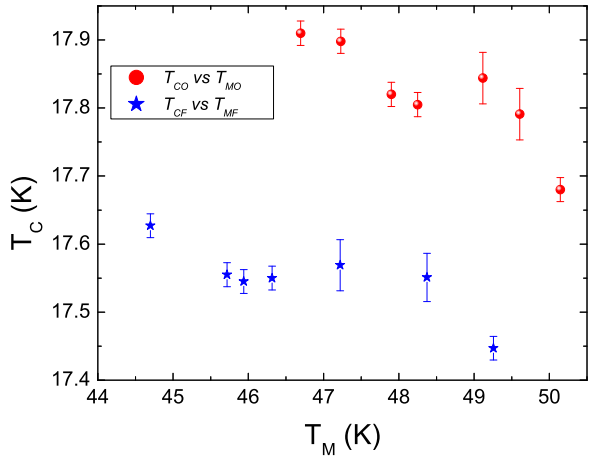}
\end{center}
\caption{(Color online) Plots for the transition temperatures;
onsets, $T_{CO}$ and $T_{MO}$ and finals, $T_{CF}$ and  $T_{MF}$,
for both transitions and seven Nb$_3$Sn crystal samples.}
\label{fig5}
\end{figure}

In Fig. \ref{fig3}, and  \ref{fig4} can be observed slight but
discernible  differences in the onsets of the superconducting
transition temperatures, and on the sizes and widths of the
martensitic transitions. Note that there is a trend, outside of
experimental error, between the two transitions; the superconducting
temperature onset decreases as the martensitic temperature onset
increases. The differences between samples could arise, independently
from different amounts of strain or slight differences in
composition, or from an intrinsic physical property between the two
transitions, as mentioned before. The total variation of $T_C$ is
small, but quite measurable, however the total variation of $T_M$ is
large. It is perhaps not surprising that $T_C$ among the seven
crystals displays such a small variation since they were taken from
the same batch and experienced very similar growth conditions. For
the same reason, however, it is surprising that $T_M$ displays such a
large variation, namely ten times that of the $T_C$ change. In Fig.
\ref{fig5} we show the superconducting transition temperature as a
function of the martensitic transition temperature from the seven
measured samples. There, we include data of the onset temperatures,
superconducting $T_{CO}$ and martensitic $T_{MO}$, and the
superconducting $T_{CF}$ and martensitic $T_{MF}$ temperatures at the
end of the transition. A very clear correlation is observed between
T$_C$ and T$_M$. It is important to note in this figure the size of
the errors bars. In some measurements with some samples, it was quite
difficult to distinguish the onsets and finals points  of the
transition temperature on the experimental curves. So in that case
the bars are large.

The physical property that could link $T_C$ and $T_M$ is the density
of electronic states at the Fermi level. Thus, if the martensitic
transition temperature and the size of the peak anomaly decreases,
the changes will increase the density of electronic states available
for the superconducting transition. This behavior is at least
consistent with the idea of a charge density wave (CDW) associated
to the martensitic transformation as proposed initially by
Gor'kov \cite{gorkov73} and continued by Bhatt and McMillan
\cite{bhatt76} using the Landau theory. This theory presents the
possibility of a competition for the Fermi surface by the CDW and
superconductivity in order to lower the total energy of the system
by opening a CDW energy gap, and therefore using some electron
population.

\begin{figure}[btp]
\begin{center}
\includegraphics[scale=0.9]{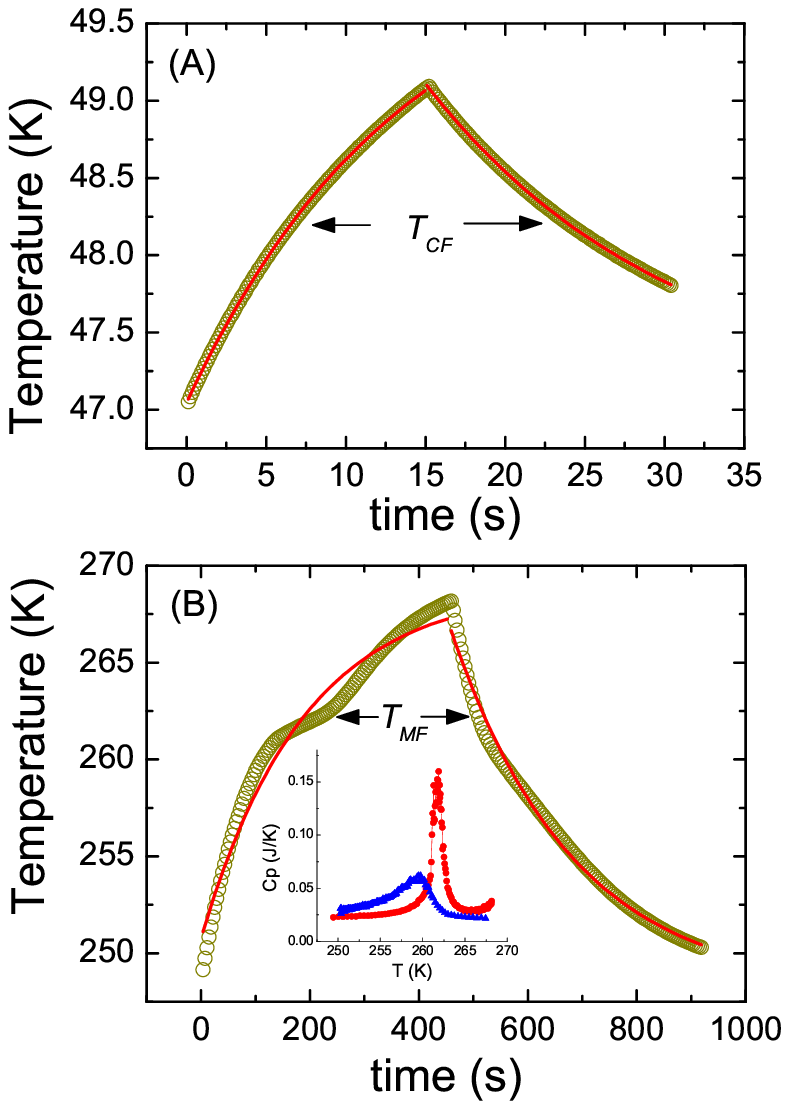}
\end{center}
\caption{(Color online) Temperature relaxation time curves in the
vicinity of the maximum of the martensitic transition of; (A) Nb$_3$Sn
and (B) Cu-Zn-Al. Note that in Cu-Zn-Al there is a distortion,
which is a characteristic of a first-order transition. Continuous
lines are an exponential decay fitting. The arrows indicate the
end of the transition temperatures. Inset of figure (B) presents
the specific heat curves in warming and cooling cycles}
\label{fig6}
\end{figure}

In order to confirm  the absence of hysteresis in the specific heat
around the martensitic transition in Nb$_3$Sn, we performed two
additional tests: the first one consisted in to measure the specific
heat of a single crystal of Cu-Zn-Al, considered by experts in the
field as a martensitic prototype alloy that shows a first order
thermodynamic transition. This was previously characterized
\cite{Tsumura88}. The results show hysteresis between the cooling and
warming cycles, in agreement with the reported austenite and
martensitic temperature values \cite{Tsumura88}. However, we have to
mention that the martensitic anomaly in this compound does not show a
lambda-type transition, as measured with our PPMS. The reason is due
to the algorithm used for the calculation of the specific heat, which
is not adequate for a first-order transitions measurement
\cite{Lashley}. The second test was to calculate the specific heat
from the raw temperature-time data using the Bachmann`s approach, as
 suggested by Lashley et al. \cite{Lashley}. The results obtained
with the Cu-Zn-Al sample shows an hysteresis and the transition
anomaly four times higher than the observed with the PPMS. The same
approach was applied to the Nb$_3$Sn crystal samples, the results
show practically the same curves obtained by the PPMS algorithm and
with the Bachmann's approach, the difference consists in that the
specific heat determined with the Bachmann's approach is noisy.

\begin{figure}[btp]
\begin{center}
\includegraphics[scale=0.95]{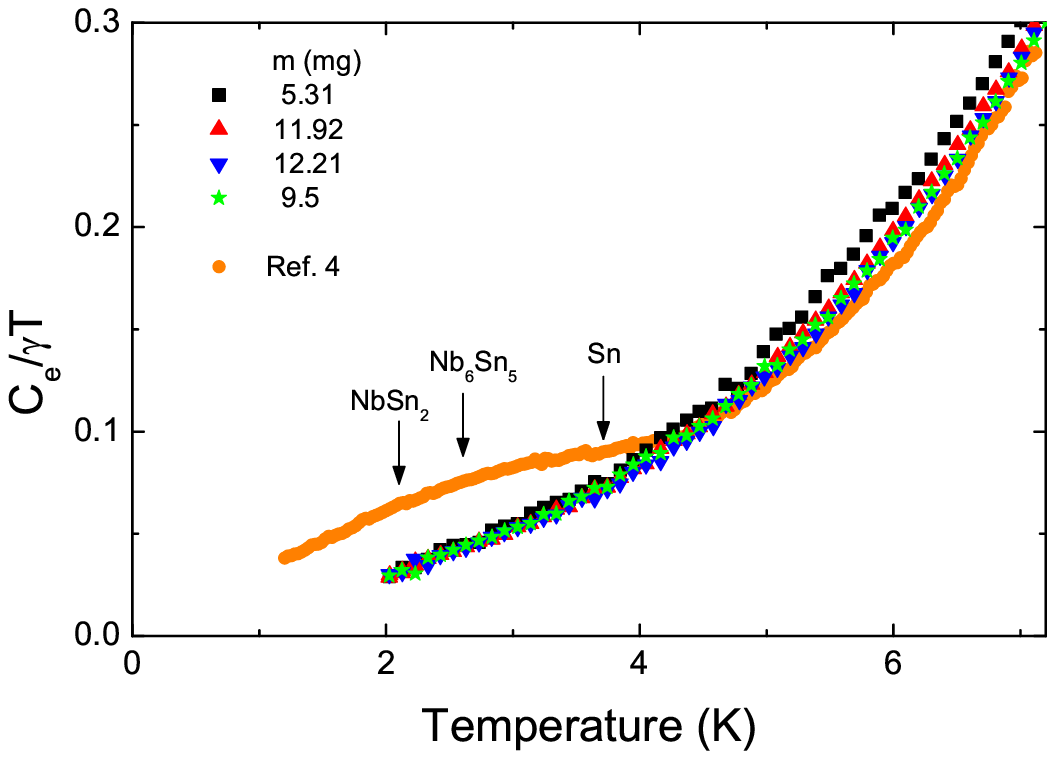}
\end{center}
\caption{(Color online) Specific heat measurements of four of our
single crystals specimens of Nb$_3$Sn at low temperature, and
specific heat measurements by authors of Ref. 4. Note the feature
below 4 K that the authors of this reference claim is related to a
second energy gap. Our measurements in pure single crystals do not
show that anomaly. Arrows indicate the T$_C$
of possible superconductors that may be formed in plycrystalline
samples}.
\label{fig7}
\end{figure}

In Fig. \ref{fig6} we show the thermal relaxation data for the
Nb$_3$Sn crystal sample (A) with mass equal to 6.5 mg, and for the
Cu-Zn-Al martensitic single crystal sample (B). This figure
illustrates the difference between a first and second order
transition. There we indicate with arrows the temperature of the end
of the transition, $T_{MF}$, were the change in the specific heat is
most notorious. The continuous line is a fit based on the sum  of two
exponential decay functions, that involve two relaxation time
constants. Note that the fitting reproduces adequately the Nb$_3$Sn
relaxation experimental data. In Fig. 6(B) we note that the fitting
to the experimental relaxation curves does not reproduce the Cu-Zn-Al
relaxation experimental data. In the inset of figure 6(B) we present
the specific heat curves of the Al-Cu-Al alloy measured in the two
cycles; lowering and rising the temperature. A  hysteresis is clearly
noted. The features are similar to early specific heat studies and
measurements performed by Tsumura et al. \cite{Tsumura88}.

Fig. \ref{fig7} shows an interesting comparison of the normalized
specific heat of four of our specimens and Guritanu et al., early
measurements in their polycrystalline sample. This plot is displayed
in term of C/$\gamma$T versus temperature. At the
temperature below 6 K it seems that our samples do not present any
anomaly as  Guritanu, et al. do, and that they mention as the feature
of a second energy gap. In this figure we included  the
superconducting transition temperatures of impurities that frequently
are found in Nb$_3$Sn polycrystalline samples (indicated by arrows)
of possible contaminants. These common impurities are  NbSn$_2$
(T$_C$=2.1 K), Nb$_6$Sn$_5$ (T$_C$=2.6 K) and Sn (T$_C$= 3.7 K). Note
that these impurities have transition temperatures in the temperature
range where the C$_p$ reported in Ref. 4 presents the anomaly, and
which is absent in our measurements.

\section{Conclusions}

In conclusion, in this work we present new specific heat measurements
performed in high quality Nb$_3$Sn single crystals. Due to the
excellent quality of the samples, we have observed a clear specific
heat feature at the martensitic transition. In multiple cooling and
heating cycles, we did not observe hysteretic behavior in the
specific heat measurements as function of temperature. Our
measurements imply that the so-called martensitic transition is a
second order or weakly first order transition, perhaps related to a
charge density wave or Peierls distortion. The  samples SN1 and SN4
examined by X-ray diffraction  present the specific heat anomaly in
very different form: whereas SN1 has a very well defined sharp peak,
with onset  at about 50 K, SN4 presents only a small feature with
onset at about 46 K. We speculated that this effect is a consequence
of vacancies in the Nb chains, and consequently the martensitic
anomaly could be  related to a Peierls distortion.

This study shows the existence of a correlation between the
superconducting transition temperature and the anomaly at high
temperature.

At lowest temperatures, from around 2 - 20 K, we observed only one
superconducting energy gap feature. Our data of the specific heat
measurements  in this temperature interval, do not show any feature
or anomaly at low temperature that could be related to a second
energy gap.

\begin{acknowledgments}
We thank R. Black from Quantum Design for help with the new software
and helpful discussions. To G. W Webb from University of California, for
kindly provide the samples used for this study. R. Escudero. thanks support of DGAPA-UNAM. We also
thank to F. Silvar for Helium provisions.
\end{acknowledgments}

\end{document}